\documentclass[twocolumn,aps,prb,showpacs]{revtex4}
\usepackage{epsfig}

\begin{document}


\title{Nonmagnetic impurity resonance states as a test of superconducting pairing symmetry in CeCoIn$_{5}$}
\author{Bin Liu}
\email{liubin@bjtu.edu.cn}

\affiliation{Department of Physics, Beijing Jiaotong University, Beijing 100044, China}

\begin{abstract}

We theoretically study the effect of a nonmagnetic impurity in heavy fermion superconductor CeCoIn$_{5}$ within a coherent three-dimensional Anderson lattice model and the T-matrix approximation approach. By considering two known possible pairing symmetry candidates $d_{x^{2}-y^{2}}$ and $d_{xy}$, we find that although both total density of states exhibit a similar V-shaped gaplike feature, only $d_{x^{2}-y^{2}}$-wave pairing symmetry gives rise to robust intragap impurity resonance states reflected by the resonance peaks near the Fermi energy in the local density of states. These features can be readily probed by scanning tunneling microscopy experiments, and are proposed to shed light on the pairing symmetry and provide hints on the microscopic mechanism of unconventional superconductivity in the Ce-based heavy fermion superconductors.

\end{abstract}
\pacs{74.25.Jb, 74.20.Pq, 74.50.+r, 74.62.En}

\maketitle

Recently, the interplay of antiferromagnetic (AF) order and unconventional superconductivity in Ce-based heavy fermion superconductors CeMIn$_{5}$ (M = Co, Rh, Ir) have been intensively studied \cite{Petrovic,Hegger,Fisher,Kawasaki,Yashima,Zheng,Jaime,Kawasaki0,Sarrao,Stock,Kenzelmann,Nair,Fisk}. For instance, CeCoIn$_{5}$ is a superconductor with the highest transition temperature T$_{c}$$\approx2.3K$ whereas CeRhIn$_{5}$ orders antiferromagnetically below T$_{N}$$\approx3.7K$\cite{Sarrao}. On the other hand, superconductivity is observed in the latter compound by application of pressure whereas unconventional superconductivity in CeCoIn$_{5}$\cite{Sarrao} emerges in close proximity to an AF quantum critical point as in the cuprates and pnictides superconductors. Moreover, neutron scattering experiments indicate strong AF quasielastic excitations at wavevectors Q=($\frac{1}{2}\frac{1}{2}\frac{1}{2}$) and equivalent positions in the paramagnetic regime. When entering the superconducting state, the magnetic excitations spectra by inelastic neutron scattering show the appearance of a sharp spin resonance\cite{Stock}. These finding underline the analogy to the cuprate high-temperature superconductors\cite{Sidis,Hayden} and the new iron superconductors\cite{Zhang,Dai}, where AF spin fluctuations may actually mediate unconventional superconductivity.

So far, the superconducting pairing symmetry in CeCoIn$_{5}$ has been discussed from both experimental and theoretical sides, it has not yet been determined unambiguously. Soon after the discovery of CeCoIn$_{5}$ material, its Fermi surface (FS) has been studied in detail by quantum oscillation, which consists of nearly cylindrical one and small ellipsoidal ones. The cylindrical sheets reflect quasi-two-dimensional(2D) character, by analogy with cuprates. Then the pairing state in CeCoIn5 has been widely believed to be unconventional with d-wave symmetry with vertical line node. The early thermal conductivity and specific heat have been measured in a rotating magnetic field, and gave a controversial result on whether CeCoIn$_{5}$ has a superconducting gap with $d_{x^{2}-y^{2}}$ or $d_{xy}$ pairing symmetry\cite{Matsuda,Matsuda0,Izawa,Aoki}. The latter $d_{xy}$ pairing symmetry also inferred from the anisotropy in the high-field superconducting phase\cite{Ikeda}. Recent specific heat measurements\cite{An} from the same group of Ref. 21 observed the predicted inversion of the oscillations\cite{Vekhter} at lower temperature, which seemed to solve the dispute in favor of $d_{x^{2}-y^{2}}$ case. In addition,
detection of a magnetic resonance in neutron scattering experiment\cite{Stock} and Bogoliubov quasiparticle scattering interference imaging technique also suggest $d_{x^{2}-y^{2}}$ pairing symmetry may be more favorable\cite{Allan}. Theoretically, detailed calculations of the spin resonance show that the resonance can appear only for the $d_{x^{2}-y^{2}}$ pairing symmetry but not in the $d_{xy}$ case\cite{Eremin}. The recent Field-angle-resolved anisotropy in the specific heat calculations also indicates the different features by considering the pairing gap function $d_{x^{2}-y^{2}}$ and $d_{xy}$\cite{Das}.

Although the ideally field-angle resolved thermal conductivity and specific heat measurements can give the position of the nodes, they rely on the ability to accurately model the true electronic structure, which in fact is poorly understood in heavy fermion materials. It is also interesting to find that the universal limit of the residual term in the thermal conductivity is not obeyed with the La-doped in CeCoIn$_{5}$\cite{Tanatar}, where the contrasting behavior between thermal conductivity and specific heat with increasing impurities reveals the presence of uncondensed electrons coexisting with nodal quasiparticles. The recent Muon knight shift measurements also found that the magnetic field dependence of the reduction of the muon knight shift is not in proportion to $\sqrt{H}$, which is roughly explained by the Fermi liquid relation and is inconsistent with simple expectation for a d-wave superconductors\cite{Haga1}. These facts likely reflect the multiband nature of superconductivity in CeCoIn$_{5}$, similar to the story of Fe-based superconductors\cite{Zhang,Dai}.

On the other hand, de Hass-van Alphen (dHvA) experiments in CeCoIn$_{5}$ clearly indicate the FS is three dimensional (3D)\cite{Haga,Hall,Shishido}. If quasi-2D FS (a large FS denoted as $\beta$-band in the dHvA experiments) is strictly cylindrical along the $k_{z}$ direction, the hot lines would be parallel to $k_{z}$. In this case, a neutron scattering resonance could be interpreted as a 2D spin excitation by analogy with cuprates as an evidence of $d_{x^{2}-y^{2}}$ pairing symmetry with vertical line node, and should be observed for the whole set of momenta Q=($\frac{1}{2}\frac{1}{2}x$) with $0\leq x\leq\frac{1}{2}$. However this explanation of the resonance in CeCoIn$_{5}$ disagrees with the experiments where the neutron scattering resonance is only found at Q=($\frac{1}{2}\frac{1}{2}\frac{1}{2}$)\cite{Stock,Eremin}. In Ref. 33 the authors argued that the absence of strong resonances at other momenta may due to the facts that the quasi-2D FS was not a perfect cylindrical as evidences by the existence of three different dHvA orbits, that gap parameter generally varied along the z axis, and that 3D FS should also had contributions\cite{Chubukov}. Thus they proposed a potential candidate-the "magnon" scenario for spin resonance in 3D superconductor in CeCoIn5, which didn't require a $d_{x^{2}-y^{2}}$ gap.

Since most of the experimental evidence for d-wave pairing symmetry is indirect, further theoretical and direct experimental work such as ARPES measurements and phase sensitive experiments is still needed and necessary to identify the order parameter in CeCoIn$_{5}$. In this paper, we propose to use local electronic structure around a single nonmagnetic impurity to probe the pairing symmetry in CeCoIn$_{5}$ superconductor, since such properties have proved to be successful in identifying the unconventional pairing states of different classes of superconductors\cite{Zhu,Zhu1,Onari,Zhang1,Dubi,Liu,Liu0}. Within a coherent 3D Anderson lattice model and the T-matrix approximation approach, we theoretically calculate the local density of states in the unitary limit of impurity scattering, and find that although both total density of states exhibit a similar V-shaped gaplike feature, the impurity induced intragap resonance state only occurs when the pairing symmetry is of $d_{x^{2}-y^{2}}$, similar to the case of $d$-wave cuprate superconductor. Our prediction can be directly measured by scanning tunneling microscopy experiments in heavy fermion superconductor CeCoIn$_{5}$

According to band structure calculations\cite{Maehira,Onuki,Haga,Hall,Shishido}, CeCoIn$_{5}$ comprises several $f$-bands and conduction bands which are hybridized in a complex manner. Due to a large spin-orbit coupling the Ce-4f electron states are split into upper j=7/2 and lower j=5/2 states, the latter one is further split into three crystalline electric field (CEF) Kramers doublet states. Because CEF splitting energy is much bigger than the heavy quasiparticle band width (about 4 meV), we can restrict to the lowest CEF doublet which has an effective pseudo-spin 1/2. Thus we here consider a 3D coherent Anderson lattice model which could reproduce the result of band calculations, realize real FS, and is also technically manageable for a T-matrix calculation in the superconducting states. In fact, this has been done previously for CeCoIn$_{5}$\cite{Tanaka}, and the resulting hybridized quasiparticle energy dispersion can be read as:
\begin{eqnarray}
 E_{\bf k\pm}&=&\frac{1}{2}[(\varepsilon^{c}_{\bf k}+E^{f}_{\bf k})\pm\sqrt{(E^{f}_{\bf k}-\varepsilon^{c}_{\bf k})^{2}+4V^{2}_{\bf k}}]
\end{eqnarray}
where $\varepsilon^{c}_{\bf k}$ and $E^{f}_{\bf k}$ are the effective $f$-band and the conduction band dispersions, respectively, and $V_{\bf \bf k}$ is the effective hybridization strength, which is renormalized by the on-site $f-f$ Coulomb repulsion. The detailed $\varepsilon^{c}_{\bf k}$, $E^{f}_{\bf k}$ and $V_{\bf k}$ as well as the parameters are defined as
%
in Ref. 43, and the resulting FS is shown in Fig. 1. Note that band $E_{\bf k-}$ reproduces the above FS and is denoted as the $\beta$-band in the de Hass van Alphen experiments\cite{Haga,Hall,Shishido}. While band $E_{\bf k+}$ remains above the Fermi energy, and thus has no contribution to the FS topological structure. Therefore, the novel low energy electronic state properties of CeCoIn$_{5}$ mainly originate from the band $E_{\bf k-}$.
\begin{figure}[tbp]
\begin{center}
\includegraphics[width=0.8\linewidth]{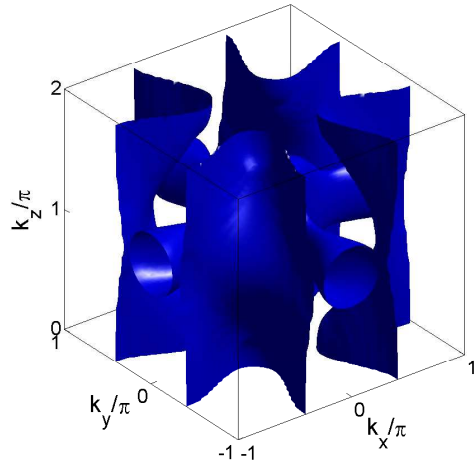}
\end{center}
\caption{(Color online) Calculated Fermi surface for CeCoIn$_{5}$ using the band structure parameters defined in Ref. 42.}
\end{figure}

In the superconducting state, the bare Green's function within the Nabu space is given by
\begin{eqnarray}
\hat{G}^{-1}_{0}({\bf
k}, i\omega_{n})=i\omega_{n}\hat{1}-\left (\matrix{E_{\bf k+} &\Delta_{\bf k}
&0 &0\cr \Delta_{\bf k} &-E_{\bf k+}
&0 &0\cr 0 &0 &E_{\bf k-} &\Delta_{\bf k}\cr 0 &0 &\Delta_{\bf k} &-E_{\bf k-}\cr}\right)
\end{eqnarray}
where $\omega_{n}=(2n+1)\pi T$ is the Matsubara frequency for fermions. The superconducting gap function is described by $\Delta_{\bf k}$.
Then the corresponding bare real-space Green's function can be obtained from the Fourier transform as
\begin{eqnarray}
\hat{G}_{0}(i,j;i\omega_{n})=\frac{1}{N}\sum_{\bf k}e^{i\bf k\cdot\bf
R_{ij}}\hat{G}_{0}({\bf k},i\omega_{n}),
\end{eqnarray}
where $\bf R_{ij}=\bf R_{i}-\bf R_{j}$ with $\bf R_{i}$ being
lattice vector and N is the number of lattice sites. In the presence of a single-site nonmagnetic
impurity of strength $U_{0}$ located at the origin $r_{i}=0$, the
site dependent Green's function in term of the T-matrix
approach can be obtained as
\begin{eqnarray}
\hat{G}(i,j;i\omega_{n})&=&\hat{G}_{0}(i,j;i\omega_{n})\nonumber\\&+&\hat{G}_{0}(i,0;i\omega_{n})\hat{T}(i\omega_{n})\hat{G}_{0}(0,j;i\omega_{n}),
\end{eqnarray}
where
\begin{eqnarray}
\hat{T}(i\omega_{n})=\frac{\hat{U}_{0}}{\hat{1}-\hat{G}_{0}(0,0;i\omega_{n})\hat{U}_{0}}
\end{eqnarray}
and the potential scattering matrix takes the following structure:
\begin{eqnarray}
\hat{U}_{0}=\left (\matrix{U_{0} &0
&V &0\cr 0 &-U_{0}
&0 &-V\cr V &0 &U_{0} &0\cr 0 &-V &0 &-U_{0}\cr}\right)
\end{eqnarray}
where $U_{0}$ and $V$ are the strength of the intra- and interband scattering potential.

The local density of states (LDOS) which is proportional to the local differential tunneling conductance measured by STM experiment can be expressed as:
\begin{eqnarray}
\rho(i,\omega)=-\frac{1}{\pi}{\rm Im} {\rm Tr} [\hat{G}(i,i;i\omega_{n}\rightarrow\omega+i0^{\dagger})]
\end{eqnarray}
The above scheme is sufficiently general to capture the essential properties of the single impurity scattering in a two-band superconductor. For the present case, the FS crossing originates only from band $E_{\bf k-}$ as discussed above, while band $E_{\bf k+}$ contributes little to the density of states (DOS) near Fermi energy. Therefore, it will be reasonable in the following calculations to only consider the impurity scattering effect in intraband $E_{\bf k-}$, and ignore the scattering from band $E_{\bf k+}$ and the interband impurity scattering (i.e., $V=0$).

Before investigating the effect of the single impurity scattering, we need to firstly look into the properties of DOS. Notice that the corrugated FS of CeCoIn$_{5}$ is characterized by three-dimensionality and is not a perfect cylindrical alone the (0 0 1) line, following the method applied in Ref. 44 we firstly restrict to the $ab$ plane by averaging over the momenta in the $k_{z}$ direction, and analyze the DOS and local electronic structure induced by a nonmagnetic impurity for each slice of the FS at a particular $k_{z}$, and then by averaging over the individual DOS and LDOS  of each $k_{z}$ slice to obtain the final total DOS and LDOS along the (001) direction.

The considered pairing symmetry includes two possible candidates as discussed above, namely $d_{x^{2}-y^{2}}$ gap symmetry with
\begin{eqnarray}
\Delta_{\bf k}=\Delta_{0}(cosk_{x}-cosk_{y})/2
\end{eqnarray}
and $d_{xy}$ gap symmetry with
\begin{eqnarray}
\Delta_{\bf k}=\Delta_{0}(sink_{x}sink_{y}).
\end{eqnarray}
The magnitude of the d-wave gap parameter $\Delta_{0}$ should in principle be determined self-consistently, but for sake of allowing for an analytic calculation, it is reasonable to assume its value is known. And also for the convenience of comparison, we assume the value $\Delta_{0}$ is the same for different $k_{z}$ layer and for both two pairing symmetries.

We now turn to analyze the FS topological structure and the gap function in the first Brillouin zone for individual $k_{z}$ as shown in Fig. 2. The upper panels of Fig. 2 represent the FS evolution with the $d_{x^{2}-y^{2}}$ gap function, where the node lines cut the FS (red solid line) at any value of $k_{z}$. While in the lower panels, the properties of the nodal structure is completely \textit{different}. For small values $k_{z}<0.6\pi$ the gap function with $d_{xy}$ symmetry has no node points on the FS, but changes sign across two neighboring FS arc. With increasing $k_{z}$ to about $0.6\pi$ value the inner FS appears and gives rise to the nodal structure as seen in Fig. 2e. As $k_{z}$ gradually increases and approaches $\pi$ in Fig. 2f, a new FS topological structure occurs, then the node line keeps away from the FS and the node points on the FS again disappear.

\begin{figure}[tbp]
\begin{center}
\includegraphics[width=1.0\linewidth]{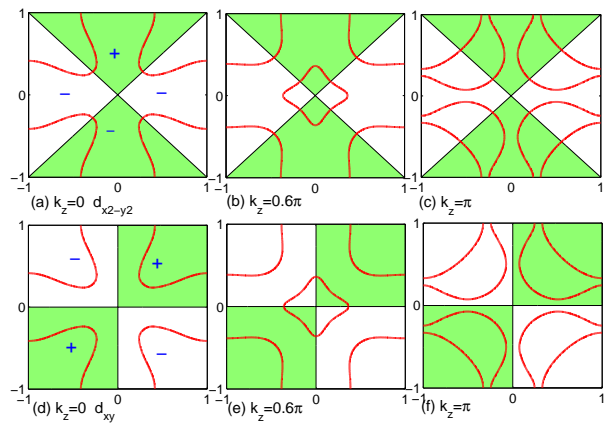}
\end{center}
\caption{(Color online) The Fermi surface topological structure and the pairing gap function in the first Brillouin zone for individual $k_{z}$. The upper panels indicate the pairing state with $\Delta_{\bf k}=\Delta_{0}(cosk_{x}-cosk_{y})/2$ for a) $k_{z}=0$, b) $k_{z}=0.6\pi$, and c) $k_{z}=\pi$, while the lower panels is for the case of $d_{xy}$ gap symmetry with  d) $k_{z}=0$, e) $k_{z}=0.6\pi$, and f) $k_{z}=\pi$. The red solid lines denote the FS and the black lines indicate the node lines, $\pm$ denote the sign of the superconducting gap. }
\end{figure}

The effect of the nodal structure on the FS can be clearly reflected by the calculated DOS which is proportional to the differential tunneling conductance tested by Tunneling experiment. In Fig. 3, the DOS for the normal state (black solid line), superconducting state with $d_{x^{2}-y^{2}}$ (red dashed line) and $d_{xy}$ (blue dotted line) gap symmetry are plotted for different cut of the FS at $k_{z}=0,0.6\pi,\pi$. For the $d_{xy}$ pairing symmetry, a V-shaped DOS only at $k_{z}=0.6\pi$ as shown in Fig. 3b is exhibited reflecting the existence of nodal structure, while in other values of $k_{z}$ the DOS is characterized by a U-shaped feature due to the nodeless gap structure and is very similar to the case of conventional s-wave superconductors. While for the $d_{x^{2}-y^{2}}$ case, the DOS always behaves to be V-shaped character at all value of $k_{z}$ because of the sign change within each FS arc. We also find that the DOS in normal state at small values of $k_{z}$ is rather smaller near the Fermi energy compared to the case of larger $k_{z}\subseteq[0.6\pi,\pi]$.
\begin{figure}[tbp]
\begin{center}
\includegraphics[width=1.0\linewidth]{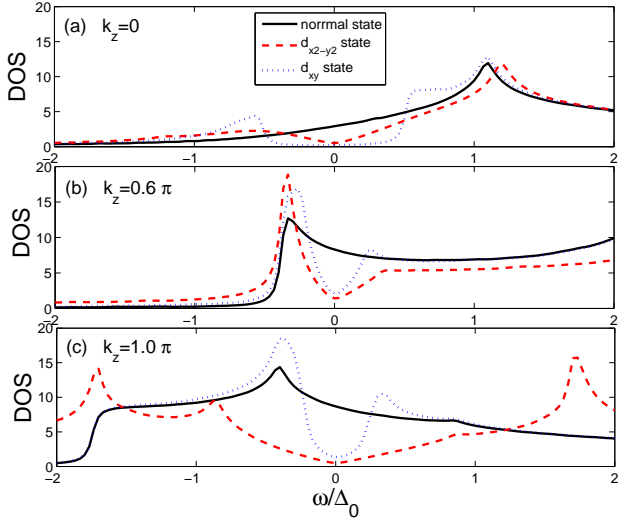}
\end{center}
\caption{(Color online) The DOS as a function of energy $\omega/\Delta_{0}$ for individual $k_{z}$ with a) $k_{z}=0$, b) $k_{z}=0.6\pi$, and c) $k_{z}=\pi$. The black solid line indicates the normal state, and the red dashed line and blue dotted line denote the superconducting states with $d_{x^{2}-y^{2}}$ and $d_{xy}$ gap symmetry, respectively.}
\end{figure}

We now proceed to analyze the response of the local electronic structure to the single nonmagnetic impurity scattering in the superconducting state of CeCoIn$_{5}$. In Fig. 4 we plot the LDOS of quasiparticles on the impurity's nearest neighboring site for different FS cuts at $k_{z}=0,0.6\pi,\pi$ considering the impurity scattering strengthen in the unitary limit $U_{0}=100t$. The LDOS at $U_{0}=0$ (black solid line) which is equivalent to DOS in the clean system, shows two coherent peaks with different spectral weight due to the particle-hole asymmetry near the gap edges and other van Hove singularity peaks originated from the particular FS topological structure at different value of $k_{z}$.
\begin{figure}[tbp]
\begin{center}
\includegraphics[width=1.0\linewidth]{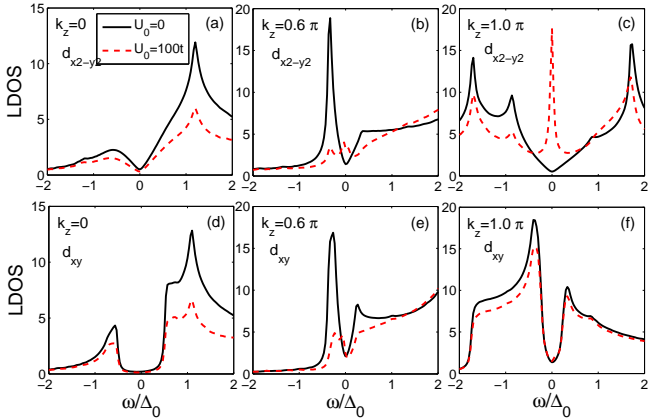}
\end{center}
\caption{(Color online) The LDOS spectra of quasiparticles on the impurity's nearest neighboring site for different scattering potentials $U_{0}$. The upper panels indicate the pairing state with $\Delta_{\bf k}=\Delta_{0}(cosk_{x}-cosk_{y})/2$ for a) $k_{z}=0$, b) $k_{z}=0.6\pi$, and c) $k_{z}=\pi$, while the lower panels is for the case of $d_{xy}$ gap symmetry with  d) $k_{z}=0$, e) $k_{z}=0.6\pi$, and f) $k_{z}=\pi$.}
\end{figure}

For the superconducting state with $d_{x^{2}-y^{2}}$ pairing symmetry as shown in the upper panels of Fig. 4a-4c, we find that impurity induced resonance states near the Fermi energy occurs denoted by the resonance peaks (red dashed line), which is the result of the sign change within each FS arc due to the node line cutting the FS, and is similar to what happens in unconventional cuprate superconductors. At the same time, the superconducting coherent peaks are heavily suppressed. We also notice that the spectral weight of the resonance peak strongly depends on the special FS topological feature and therein the nodal structure. The impurity induced intragap resonance peak is almost invisible in Fig. 4a, then is enhanced in Fig. 4b, and finally is very sharp at the zero energy in Fig. 4c. This is because that the DOS contributions of the scattering electrons increase from the $k_{z}=0$ to $k_{z}=\pi$ due to the special nodal structure on the FS as shown in Fig. 2a-2c, the spectral weight of the resonance peak is thus correspondingly enhanced. While for the $d_{xy}$ pairing symmetry case , we find that the impurity induced intragap resonance peak only occurs at $k_{z}=0.6\pi$ as shown in Fig. 4e, while at other value of $k_{z}$ disappears and is replaced by resonance peaks near the gap edges. This is consistent with the aforementioned nodal structure and DOS.

\begin{figure}[tbp]
\begin{center}
\includegraphics[width=1.0\linewidth]{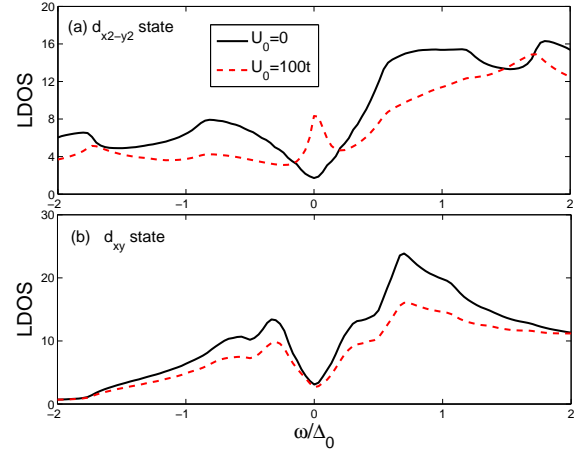}
\end{center}
\caption{(Color online) The calculated total LDOS spectra averaged in the (001) direction on the impurity's nearest neighboring site for different scattering potentials $U_{0}$ with a) $d_{x^{2}-y^{2}}$ pairing state and b) $d_{xy}$ pairing state.}
\end{figure}
Since the observable local electronic structure in CeCoIn$_{5}$ is of the three-dimensional (3D) feature and should be an average over the FS slices at different $k_{z}$, we have to consider the calculated DOS and LDOS averaged in the (001) direction. In this case, we plot the averaged DOS and LDOS for the superconducting states with $d_{x^{2}-y^{2}}$ pairing symmetry and $d_{xy}$ pairing symmetry in Fig. 5. We find that although both total DOS exhibits a similar V-shaped gaplike feature, only the $d_{x^{2}-y^{2}}$-wave pairing symmetry gives rise to robust impurity resonance states reflected by a zero energy resonance peak in LDOS. Our result confirms the recent experimental results where such pairing breaking effect by impurities has been indirectly measured in CeCoIn$_{5}$ after doping hole or electron\cite{Fisk,Tanatar}. We propose that these features can be directly measured by scanning tunneling microscopy (STM) experiments, and then shed light on the pairing symmetry and pairing mechanism in CeCoIn$_{5}$ and other Ce-based heavy fermion superconductors, since they have the similar HoCoGa$_{5}$-type electronic structure.

In conclusion, by applying the T-matrix approximation approach we have studied the effect of a single nonmagnetic impurity in CeCoIn$_{5}$ superconductor within a coherent Anderson lattice model which can reproduce the real 3D FS topological feature. We have found that, considering two types of pairing symmetry $d_{x^{2}-y^{2}}$ and $d_{xy}$, only $d_{x^{2}-y^{2}}$ pairing gives rise to robust intragap impurity induced resonance state near the Fermi energy in the unitary limit of impurity scattering, though both pairing gap in the superconducting state indicate the similar V-shaped feature of DOS. Based on these results, we propose to use STM experiment to test the local electronic structure around nonmagnetic impurities so as to identify the pairing symmetry and provide hints on the microscopic mechanism of unconventional superconductivity in the Ce-based heavy fermion superconductors.

After completing the present work, we are aware of the recent high-resolution STM experiment on the CeCoIn$_{5}$ superconductor\cite{Zhou}, where due to the cleaving procedure which could cut the surface at different $k_{z}$ points, the STM spectrum is available for different $k_{z}$ plane. After analyzing $k_{z}$ plane measured in above experiment and its FS topology, we find it is basically located in the $k_{z}\subseteq[0.6\pi,\pi]$ as shown in Figs. 2b-2c and Figs. 2e-2f, and impurity-induced intragap bound states experimentally probed indeed confirm our theoretical predictions as seen in Figs. 4b-4c for the $d_{x^{2}-y^{2}}$ pairing state.

We acknowledges helpful discussions with Prof. Shiping Feng, Ilya Eremin, and Zhi Wang. This work was supported by the National Natural Science Foundation of China (NSFC) under Grants No. 11104011, Doctoral Fund of Ministry of Education of China under Grants No. 20110009120024, Research Funds of Beijing Jiaotong University under Grant No. 2013JBM092, and The Project sponsored by SRF for ROCS, SEM.

\end{document}